\documentstyle[version2,aps]{revtex}
\begin{document}
\draft
\columnsep -.375in
\twocolumn[
\begin{title}
Landauer Conductance of Luttinger Liquids
with Leads
\end{title}
\author{Dmitrii L. Maslov$^{(a,b,c)}$  and Michael
Stone$^{(a)}$}
\begin{instit}
$^{(a)}$Department of Physics and $^{(b)}$Materials
Research
Laboratory\\
University of Illinois at Urbana-Champaign
Urbana, IL 61801-3080, USA\\
$^{(c)}$Institute for Microelectronics Technology,
Academy of Sciences of Russia, Chernogolovka, 142432
Russia
\end{instit}
\begin{abstract}
 We show that the dc conductance of a quantum wire
containing a Luttinger liquid and attached to
non-interacting leads is
given by $e^2/h$ per spin orientation, regardless of the
interactions in
the wire. This explains the recent observations of the
absence of
conductance renormalization in long high-mobility $GaAs$
wires by
Tarucha, Honda and Saku (Solid State Communications {\bf
94}, 413
(1995)).
\end{abstract}
\pacs{ PACS numbers: 72.10.Bg, 73.20.Dx
\hfil\break
Preprint number: P-95-05-023-i}
]
\narrowtext

For non-interacting electrons, the conductance of narrow
ballistic
 quantum wires connected to wide reservoirs is quantized
in units of
$e^2/h$ \cite{wees,wharam}.  When the effects of
interactions are
included this result is expected to be modified. In
particular, when the
electrons in the wire form a one-dimensional
Luttinger liquid \cite{haldane}, the
conductance is believed to be $Ke^2/h$ per spin
orientation
\cite{apel,fisher-prb,fukuyama}, where $K$ is the
interaction dependent
parameter characterizing the Luttinger liquid.  For
non-interacting
electrons $K=1$. For repulsive interactions $K<1$, and
the conductance
should be reduced.

A recent experiment on very long $GaAs$ high mobility
quantum wires
\cite{tarucha} casts doubt on this picture.  It is known
that the same
parameter $K$ enters the temperature dependence of the
impurity
correction to the conductance, and using this the
authors of
\cite{tarucha} where able to estimate $K$ to be about
$0.7$ for the
electron gas in their wires, implying a conductance
reduction of $30\%$
in the ballistic limit. The actual reductions observed
are only a few
percent of $e^2/h$ however.

This observation leads us to re-assess the conventional
analysis of the
charge transport in Luttinger wires. In this paper we
will argue that
the conductance of a quantum wire attached to
one-dimensional
non-interacting leads (which are intended to  model the
higher-dimensional Fermi-liquid reservoirs) is $e^2/h$
regardless of
the interactions in the wire itself. This is because the
finite
resistance of a ballistic wire is a {\it contact
resistance}
\cite{imry,landauer,glazman,levinson} and comes entirely
from processes
that take place outside the wire, when the electrons are
not in the
Luttinger-liquid state (cf. Fig.~\ref{fig:fig1}a). When
the Umklapp scattering  is negligible (which
is the case for the low-density electron gas in
semiconductor
heterostructures), the interactions in the wire conserve
the total
momentum and thus do not change the resistance.

While it is generally asserted in the literature that
$g=Ke^2/h$ is the
correct result for interacting electrons
\cite{apel,fisher-prb,fukuyama} (at least for wires
longer than the
Fermi wavelength), we must remark there have been
previous comments
supporting the result $g=e^2/h$.  In their earlier paper
\cite{fisher-prl}, Fisher and Kane remark that the ac
conductance will
cross over to the non-interacting value at frequencies
lower than
$\omega= v_F/L$, where $v_F$ is the Fermi velocity and
$L$ is the
wire length.  Matveev and Glazman \cite{matveev} make a
similar
remark when discussing the conductance of multi-mode
interacting wires
\cite{matveev1}.

In this paper we will first calculate the conductance
for a wire containing a
homogeneous Luttinger liquid with parameter $K_W$
attached to leads
which are also homogeneous but have  parameter $K_L$. We
will see that
the conductance is given by $K_Le^2/h$. We will then
prove a general
theorem showing that even when $K$ and the density-wave
velocity $v$
vary arbitrarily in the Luttinger liquid, the
conductance is determined
by  the asymptotic value of $K$ only. As we assume that
the electrons
do not interact in the leads, this implies that $K_L=1$
and the
conductance is given by $e^2/h$.

\section{Kubo Formula Calculation}
\label{sec:Kubo}

We start with the simple case of an infinite  Luttinger
liquid (LL)
which contains three regions:  the wire ($0\leq x\leq
L$) and two leads
($x<0$
and $x>L$, respectively). The interaction parameter is
assumed to
change abruptly from $K_L$ in the leads to $K_W$ in the
wire (Fig.~\ref{fig:fig1}b). As
the
one-dimensional leads are supposed to model two- or
three-
dimensional
Fermi-liquid leads, we will put $K_L=1$ at the end of
the calculation.
In a real system with wide leads, the applied
electrostatic potential
difference $V_{-}-~V_{+}$ produces an inhomogeneous
electric field
$E(x)$ which is concentrated in the wire and decays
rapidly towards
the
open ends of the system.  In the one-dimensional model
we imitate
this
behavior by assuming that the field is zero outside the
wire. Thus the
current $I=ej$ is related to the field by
\begin{equation}
I(x,t)=\int^{L}_{0}dx'\int \frac{d\omega}{2\pi}
 e^{-i\omega t}\sigma_{\omega}(x,x'){\bar
E}_{\omega}(x'),
\label{eq:current}
\end{equation}
where ${\bar E}_\omega(x)$ is the time Fourier component
of the
electric field
and $\sigma_{\omega}(x,x')$ is the non-local ac
conductivity. In the
Matsubara
representation, $\sigma_{\omega}(x,x')$ is expressed via
the
(imaginary time) current-current correlation
function by the usual Kubo formula
\begin{equation}
\sigma_{\omega}(x,x')=-\frac{e^2}{\pi{\bar
\omega}}\int^{\beta}_{0}\langle
T_{\tau}^{*} j(x,\tau)j(x',0)\rangle e^{-i{\bar
\omega}\tau},
\label{eq:kubo}\end{equation}
where $t=i\tau$ and ${\bar \omega}$ is defined by
$\omega=i{\bar
\omega}+\epsilon$.
The (Euclidean) action of the spinless Luttinger liquid
is
\begin{equation}
S_{E}=\frac{1}{8\pi}\int \! dx
\!\int^{\beta}_{0}\!d\tau\!\frac{1}{K(x)}
\Big\{\frac{1}{v(x)}(\partial_{\tau}\phi)^2+v(x)(\partial_x\phi)^2\Big
\}.
\label{eq:acte}\end{equation}
In the bosonized form, the particle-number current is
$j=i\partial_{\tau}\phi/2\pi$,
and Eq.~(\ref{eq:kubo}) reduces to \cite{shankar}
\begin{equation}
\sigma_{\omega}(x,x')=-e^2\frac{{\bar
\omega}}{\pi}G_{{\bar
\omega}}(x,x'),
\label{eq:kubo1}\end{equation}
where
\begin{equation}
G_{{\bar
\omega}}(x,x')=\int^{\beta}_{0}\frac{d\tau}{(2\pi)^2}\langle
T_{\tau}^{*} \phi(x,\tau)\phi(x',0)\rangle e^{-i{\bar
\omega}\tau}
\label{eq:prop_def}\end{equation}
is the
propagator of the bosonic field. According to
Eq.~(\ref{eq:current}),
we need to know $G_{{\bar \omega}}(x,x')$ only for
$0\leq x'\leq L$.
The propagator
$G_{{\bar \omega}}(x,x')$ satisfies the equation
\begin{equation}
\Big\{-\partial_x(\frac{1}{K(x)}\partial_x)
+{\bar \omega}^2\Big\}G_{{\bar
\omega}}(x,x')=\delta(x-x'),
\end{equation}\label{eq:prop}
which leads to the following boundary conditions: i)
$G_{{\bar \omega}}(x,x')$
is continuous
at $x=0,L$ and $x=x'$, ii)
$\frac{1}{K(x)}\partial_xG_{{\bar \omega}}(x,x')$ is
continuous at $x=0,L$ but iii) undergoes
a jump of unit height at $x=x'$, {\it i.e.},
\begin{equation}
-\frac{1}{K(x)}\partial_xG_{{\bar
\omega}}(x,x')\Big|^{x=x'+0}_{x=x'-
0}=1.
\label{eq:jump}\end{equation}
In addition, we assume that the infinitesimal
dissipation is present in
the leads, so that $G_{{\bar \omega}}(\pm\infty,x')=0$.
(In a real-
time formulation this corresponds to outgoing wave
boundary
conditions.) The solution to the problem defined above
can be
written in the form
\begin{equation}
G_{{\bar \omega}}(x,x')=
\cases{
 Ae^{|{\bar \omega}|x}&for $x\leq 0$;\cr
 Be^{|{\bar \omega}| x}+Ce^{-|{\bar \omega}| x}&for
$0<x\leq x'$;\cr
 De^{|{\bar \omega}| x}+Ee^{-|{\bar \omega}| x}&for
$x'<x\leq L$;\cr
 Fe^{-|{\bar \omega}| x}&for $x>L$,
 }
 \label{eq:solut}
 \end{equation}
 where $A\dots F$ depend on $x'$ and ${\bar \omega}$ and
are to be
found from the
 boundary conditions. As we are interested in the
 dc limit of $\sigma_{\omega}(x,x')$, the frequency
${\bar \omega}$
can be put to zero in all
 boundary conditions except for in Eq.~(\ref{eq:jump}).
Consequently,
 $G_{{\bar \omega}}(x,x')$ becomes $x$-
and $x'$-independent in this limit and is readily
 found to be equal to $-K_{L}/2{\bar \omega}$ in all
regions. Thus
the dc limit of the non-local conductivity is
 \begin{equation}
 \lim_{{\bar \omega}\to
0+}\sigma_{\omega}(x,x')=\frac{K_Le^2}{2\pi}.
 \label{eq:dccond}\end{equation}
 For a static electric field, ${\bar
E}_{\omega}(x)=2\pi\delta(\omega)E(x)$, and
Eq.~(\ref{eq:current})
gives the $x$- and $t$-independent current
 \begin{equation}
 I=\frac
{K_Le^2}{2\pi}\int^{L}_{0}dx'E(x')=\frac{K_Le^2}{2\pi}(V
_{-}-
V_{+}), \label{eq:dccurrent}\end{equation}
 {from} which we see that the conductance is
 \begin{equation}
 g\equiv\frac{I}{V}=K_{L}\frac{e^2}{2\pi}.
\label{eq:conduct}\end{equation}
We have been using units where $\hbar=1$ or $h=2\pi$ so,
restoring
$\hbar$, we have $g=K_{L}e^2/h$. Thus the conductance is
determined by
the value of $K_L$ in the leads  and does not depend on
the value of
$K_W$ in the wire.

Notice that in this calculation we have implicitly taken
the limit
$\omega \to 0$ before the $q\to 0$ limit
\cite{lee-fisher,fenton,schulz}. The traditional order
of limits is
$q\to 0$ before $\omega \to 0$. The latter yields the
Drude formula,
which  has a divergent dc limit for perfect systems. The
former
produces a finite Landauer dc conductance even for a
perfect system.
It
furthermore corresponds to the experimental situation
where a static
field is applied over a finite region.

{}From our assumption of  Fermi-liquid leads we have
$K_L=1$,
consequently $g=e^2/h$. This result is independent of
the wire length  --- at least until the wire is so long
that impurity
scattering becomes significant.  Similar calculations
can be done for
two Luttinger-liquid wires connected in series and
separated by the
Fermi-liquid region. Again, the dc conductance is found
to be independent of
the
interactions in the wires and is given by $e^2/h$, even
in the case
when these interactions are of different strengths.

\section {Conductance Theorem}
\label{sec:theorem}

We can get some further insight into the result of the
previous
section, and establish a general theorem, by considering
a generic
inhomogeneous system and asking how it responds  when an
electric
field is switched on at some particular moment.

Let  $\Omega$ be a finite segment of the wire and assume
that the
Luttinger  liquid has values of $K$ and $v$ that vary
smoothly and
{\it
arbitrarily} within $\Omega$ but take  constant values
$K_L$, $v_L$
in
the leads outside $\Omega$.

The real time effective action for a spinless Luttinger
liquid is
\begin{equation}
S=\frac 1{8\pi}\int d^2x \frac 1{K(x)}\left\{v(x)
(\partial_x
\phi)^2-\frac 1 {v(x)} (\partial_t \phi)^2\right\}.
\label{(1.1)}
\end{equation}
We have chosen to  normalize the $\phi$ field  so that
\begin{equation}
\rho=-\frac 1{2\pi} \partial_x \phi, \qquad j=\frac
1{2\pi}\partial_t
\phi.
\label{(1.2)}
\end{equation}
The interaction with an external electromagnetic field
$A_\mu$ is
given by
\begin{equation}
S_{int}=\frac e{2\pi}\int d^2x \left\{-A_0\partial_x
\phi
+A_1\partial_t\phi\right\},
\label{(1.3)}
\end{equation}
so the  equation of motion for the field (classical or
quantum) is
\begin{equation}
\frac 1{4\pi}\left[
\partial_t\left (\frac 1 {Kv}\partial_t \phi\right)
-\partial_x\left(\frac v K \partial_x \phi\right)
\right]=\frac e{2\pi} E(x,t),
\label{(1.5)}
\end{equation}
where $E=\partial_xA_0-\partial_t A_1$ is the electric
field.

We seek a solution to this equation when the electric
field (which we
will assume to be non-zero only in the region $\Omega$)
is switched
on
in some manner at time $t=0$ and remains constant (but
not
necessarily
homogeneous) thereafter. We will select a particular
transient time
dependence for the field in a few lines, but remark here
that if the
conductance is to be well-defined then it must be
insensitive to this
choice.

We find the solution by first solving the time
independent problem
\begin{equation}
-\frac 12\partial_x\left(\frac {v(x)} {K(x)} \partial_x
\phi\right)=
eE(x).
\label{(1.6)}
\end{equation}
Call the solution to this problem $\Phi(x)$.
For example if $v=K=1$ and $E(x)=E$, $x\in [0,L]$, $=0$
otherwise,
then
\begin{equation}
\Phi(x)=
\cases{
  eEL x   &for  $ x<L$; \cr
- eE x(x-L) &for $ x\in [0,L]$;  \cr
- eEL (x-L) &for $ x>L$
}.
\label{(1.7)}
\end{equation}

In general, being the solution of a second-order
equation, the
function
$\Phi$ contains two constants of integration. One is
simply an
additive
constant and the other will be chosen to ensure, as we
have done in
Eq.~(\ref{(1.7)}),  that the outer, linear, portions of
the solution have
equal and opposite slopes. With this choice it  should
now be clear
that
\begin{equation}
\phi(x,t) = {\rm max\,} (0, \Phi(x)+\alpha t)
\label{(1.8)}
\end{equation}
will solve
\begin{equation}
\frac 1{4\pi}\left [\partial_t\left ( \frac
1{Kv}\partial_t\phi\right)-
\partial_x\left(\frac v K \partial_x \phi\right)
\right]=\frac e{2\pi} E(x)
\label{(1.9)}
\end{equation}
at late enough time, provided $\alpha$ is chosen so that
the points of
intersection of $\Phi(x)+\alpha t$ with the $x$ axis
move out in the
$\pm x$ directions at velocity $\pm v_L$. The time must
be late
enough that there are  only two such intersections (we
will call the
one on the left  $A(t)$ and on the right $B(t)$) and
they take place
outside
$\Omega$. The two points of intersection will move at
the same
speed
but in opposite directions because of the  opposite
slopes of the
linear portions of the solution.  We may find $\alpha$
by integrating
Eq.~(\ref{(1.9)}) over an interval $[a,b]$ containing
$\Omega$, but
itself
contained in $[A(t), B(t)]$.
We find
\begin{equation}
\frac 1 2 \left[\frac
{v_L}{K_L}\partial_x\phi\right]_a^b
=\int_a^b eE dx.
\label{(1.10)}
\end{equation}
The magnitude of the slopes are therefore
\begin{equation}
|\partial_x \phi|_{a,b}= \frac {K_L}{v_L} \int_a^b eE
dx.
\label{(1.11)}
\end{equation}
The requirement that the linear portions of $\phi(x,t)$
move out at
speed $v_L$ then gives
\begin{equation}
\alpha=\partial_t \phi= v_L \partial_x\phi|_b=
{K_L}\int_a^b eE dx.
\label{(1.12)}
\end{equation}
We can make this solution valid at all times by
switching on the field
in the manner given by substituting Eq.~(\ref{(1.8)}) in
Eq.~(\ref{(1.5)}).
This is the particular transient behavior for $E$
alluded above. If we
do not switch the field on in this way then there will
be a region at
the head of each outgoing wave where the solution will
be different
from Eq.~(\ref{(1.8)}), but this transient behavior will
not affect the
dc conductance.

Since the electric current $I$ is given by $I=ej=-
e\partial_t\phi/2\pi$
our solution has
\begin{equation}
I(x)=\cases{ \frac {e^2}{2\pi} K_{L}\int E(x') dx' &for
$ x \in [A(t),
B(t)]$\cr
 = 0 &   elsewhere $ $.
}
\label{(1.13)}
\end{equation}
As before, we have been working with units where
$\hbar=1$, so
again restoring $h$ in Eq.~(\ref{(1.13)}) gives
\begin{equation}
I= \frac {e^2}{h} K_L (V_{-}-V_{+})
\label{eq:(1.14)}
\end{equation}

Physically what is occuring is that once the electric
field is switched
on there are different equilibrium densities on the left
and right
sides of $\Omega$, since these regions are  at different
potentials
$V_{+,-}$. The extra charge $\delta
\rho=-\partial_x\phi/2\pi=\pm
(eK_L/\pi v_L)\delta V$ flows into and out of the
``reservoirs'' or
asymptotic leads in the form of a ``shock wave'' whose
head moves at
$\pm v_L$. This flow of charge provides the  current in
the leads
which
is obviously  equal to $I=\delta\rho v_L$. The current
is  therefore
determined only by the asymptotic values of $v$ and $K$.
As we said
above, different ways of switching on the field will
lead to transients
at the head of the wave of charge, but will not alter
the current
following the head.

The analysis of this and previous sections can readily
be generalized
for the case of  electrons with spin. In this case, the
conductance per
spin orientation is obtained from
Eqs.~(\ref{eq:dccond},\ref{eq:(1.14)}) by
replacing $K_L\to K^{\rho}_L$, where $K^{\rho}_L=1$
parameterizes the
charge part of the Luttinger liquid in the leads.

\section{Discussion}

We have shown in Secs.~\ref{sec:Kubo}, \ref{sec:theorem}
that the conductance
of a ballistic wire containing Luttinger liquid and
connected to
non-interacting leads is not renormalized by the
interactions in the
wire. It remains at the non-interacting value $g=e^2/h$
per spin orientation.

We now discuss some experimental implications of this
result.  At low
temperatures ($T<1.2$K), the temperature dependence of
the conductance
of quasi-ballistic GaAs wires  appears \cite{tarucha} to
be described
reasonably well by the theory of disordered Luttinger
liquids
\cite{apel,fukuyama}. The value of the $K$- parameter
extracted from
the exponent of this dependence is $\approx 0.7$
\cite{tarucha}. This
suggests that at higher temperatures, when the
disorder-induced
reduction of the conductance is not pronounced yet, the
conductance
should be $0.7e^2/h$, {\it i.e.}, significantly smaller
than the conductance
quantum.  The observed higher-temperature value of the
conductance is
very close to $e^2/h$ however, whereas the total change
in the
conductance in the whole range of temperatures is
$1-5\%$, depending on
the length of the wire. Although we have considered only
the case of a
pure wire in this paper, it would be reasonable to
expect that
reduction in the conductance due to disorder in the wire
would depend
on the $K$-parameter of the wire.  We thus can conclude
that the two
experimental results--the Luttinger-liquid-like
temperature dependence
at lower temperatures and absence of the conductance
renormalization at
higher temperatures--do not contradict to each other.

Finally, we note that the transport in quantum wires
in zero magnetic fields
is very different from the edge-state transport in the
fractional
quantum Hall effect system \cite{webb,moon}. In the
former case, electrons
in the wide leads are necessarily {\sl not} in the
Luttinger-liquid state and
the finite conductance $e^2/h$ arises from the
scattering of Fermi-liquid
electrons at the contacts. In the latter case, strong
magnetic field
binds electrons to the edges even in the leads
(at least in the 2DEG parts of the
leads adjacent to the constriction), thus both incoming
and
outgoing electrons are in the Luttinger-liquid state,
with the parameter $K$ given by the filling fraction
$\nu$. In this case
one expects to have  both the conductance
renormalization
in the absence of tunneling between edges ($g=\nu
e^2/h$) and the temperature-dependent
tunneling rate.

{\it Note Added:} While we were in the process of
writing this paper we
learned of a recent preprint  by Safi and Schulz (cond-
mat/9505079)
who also conclude that the conductance is not
renormalized by
the interactions in the wire.

\section{Acknowledgments}

This work was supported by the NSF under grants
 DMR94-24511 and DMR89-20538.  DLM would like to thank
Y.\ B.\ Levinson
for discussions over many years of the physics behind
the Landauer
formula.  We both thank  D.\ Loss for useful
conversations and drawing our
attention to the recent work of Safi and Schulz, as well
as to
Ref.~\cite{schulz}.  We also thank P.\ M.\ Goldbart for
his persistent
encouragement.

\figure{(a) The quantum wire containing Luttinger liquid
(LL) and connected to Fermi-liquid (FL) leads. Finite
resistance arises from
the scattering of FL electrons at the contacts. (b)
Effective 1D model
for the situation depicted in (a).
\label{fig:fig1}}
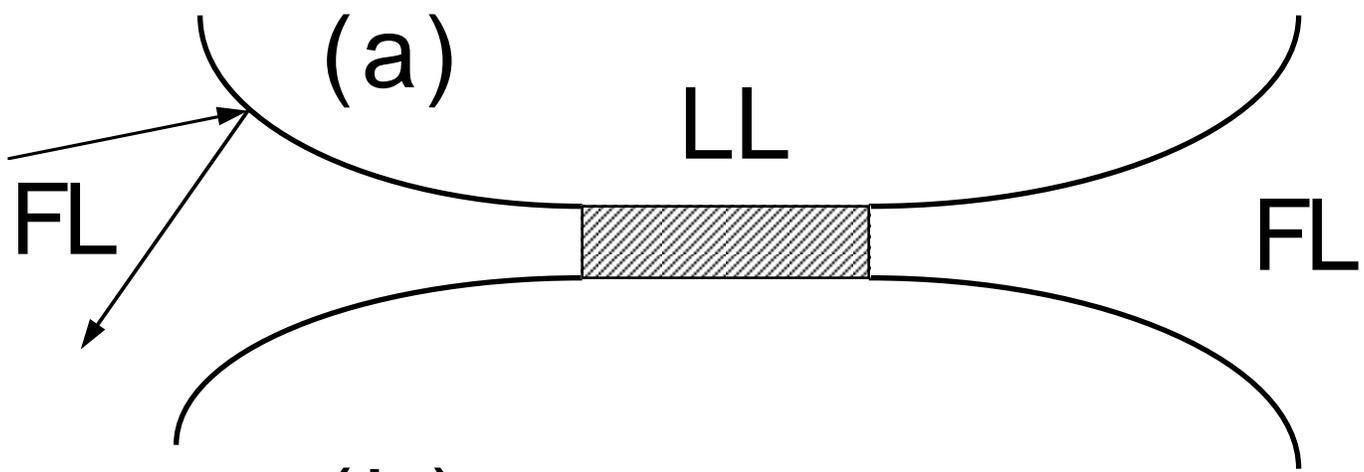
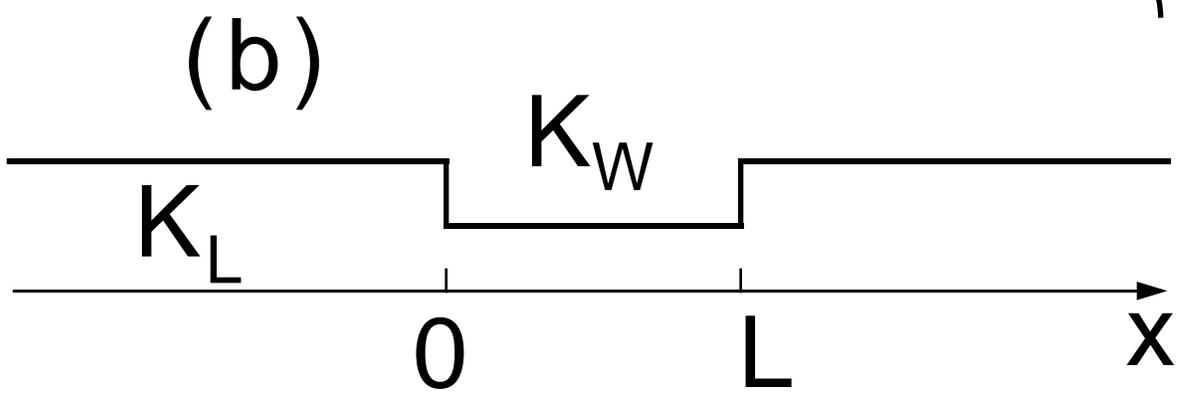
\end{document}